\documentclass[reprint,
superscriptaddress,
 amsmath,amssymb,
 aps,
pra,
]{revtex4-1}
\usepackage{dsfont}

\usepackage{nicefrac}
\usepackage{braket}
\usepackage{xcolor}
\usepackage{graphicx}
\usepackage{tikz}
\usepackage{dcolumn}
\usepackage{bm}
\usepackage{hyperref}
\usepackage[caption=false]{subfig}
\usepackage[normalem]{ulem}
\usepackage{float}
\usepackage{soulutf8}
\usepackage{algorithm}
\usepackage{algpseudocode}
\usepackage{algorithmicx}

\usepackage[hang,flushmargin]{footmisc}
\setstcolor{red}

\begin{document}

\preprint{APS/123-QED}

\title{Hidden multi-dimensional modulation side channels in quantum
protocols}

\author{Amita Gnanapandithan}
\affiliation{Department of Electrical and Computer Engineering, University of Toronto, Toronto, Canada}

\author{Li Qian}
\affiliation{Department of Electrical and Computer Engineering, University of Toronto, Toronto, Canada}
\affiliation{Center for Quantum Information and Quantum Control, University of Toronto, Toronto, Canada}

\author{Hoi-Kwong Lo}
\affiliation{Department of Electrical and Computer Engineering, University of Toronto, Toronto, Canada}
\affiliation{Center for Quantum Information and Quantum Control, University of Toronto, Toronto, Canada}
\affiliation{Quantum Bridge Technologies Inc., 108 College St., Toronto, Canada}

\date{\today}

\begin{abstract}
Quantum protocols including quantum key distribution and blind quantum computing often require the preparation of quantum states of known dimensions. Here, we show that, rather surprisingly, hidden multi-dimensional modulation is often performed by practical devices. This violates the dimensional assumption in quantum protocols, thus creating side channels and security loopholes. Our work has important impacts on the security of quantum cryptographic protocols.
\end{abstract}

\maketitle

\textit{Introduction.---}
Quantum protocols such as quantum key distribution (QKD) and blind quantum computing allow for information-theoretic, quantum-safe security \cite{XuRMP2020,Fitzsimons2017}.  Thanks to measurement-device-independent QKD (MDI QKD), all side channels on the detection end can be completely eliminated \cite{Lo2012}. Therefore, the source, as well as device-independent (DI) QKD are the last frontiers of investigation for security, with DI QKD remaining impractical due to its demanding requirements, such as near-perfect detector efficiencies \cite{Zapatero20232, XuRMP2020, Nadlinger2022,Zhang2022}. Practical implementations of non-DI protocols involve quantum state preparation that deviates from the theoretical requirements for security \cite{XuRMP2020}. The loss-tolerant protocol illustrates that as long as the prepared quantum states satisfy a dimensional assumption, minor errors will not have a significant impact on security -- otherwise, a side channel will be opened, causing a significant reduction to the maximum secure transmission distance \cite{Tamaki2014,Curras-Lorenzo2023}.

Side channels and security loopholes in general are a point of significant interest in the literature for quantum cryptographic protocols, including QKD, quantum bit commitment, quantum coin flipping and position-based quantum cryptography \cite{XuRMP2020, Xu2010, Gottesman2004, Tang2013, Zhao2008, Lydersen2010, Brassard, Jain2014}. Notable examples include the pattern effect \cite{Yoshino2018,Xing2024}, phase-remapping attack \cite{Fung2007,Xu2010}, and the laser seeding attack \cite{Sun2015}. In this Letter, we point out, rather surprisingly, that the commonly made dimensional assumption is often
violated during the modulation process in quantum state preparation. We call the resulting class of side channels ``hidden multi-dimensional modulation side channels.'' Modulation is an important part of quantum state preparation (for example, during encoding). We show that in practical experiments, the modulation process may be tied to ``hidden'' degrees of freedom, causing a violation of the dimensional assumption. As an illustrative example, we focus on the case of time-varying polarization (BB84) encoding in MDI QKD. By time-varying encoding, we are referring to encoding which varies over the duration of the quantum optical state being encoded. Here, the ``hidden'' degree of freedom is time. However, our concept is general and applies to other degrees of freedom. While we focus on QKD for illustrative purposes, we emphasize that this issue would impact quantum cryptographic protocols in general.

\textit{The Three-State Protocol.---} To understand the importance of the dimensional assumption in quantum protocols, let us start with the simple three-state QKD protocol \cite{Fung2006} consisting of the states $\ket{0} $, $\ket{1}  $, and 
$\ket{+} = { 1 \over \sqrt{2}} ( \ket{0} + \ket{1} )$. Notice that these states span two dimensions.

Practically, imperfections in preparing these states pose a security concern. Loss-tolerant QKD tells us that minor imperfections in the states do not affect the security of the protocol assuming that the states lie in a two-dimensional Hilbert space (i.e. the states satisfy the ``qubit assumption'') \cite{Tamaki2014,Pereira2019}. Such imperfections are referred to as ``qubit flaws'' \cite{Curras-Lorenzo2023}. More specifically, the QKD protocol will remain robust to channel loss (``loss-tolerant'') \cite{Tamaki2014}. However, we emphasize that this would not be the case if another dimension is present. We will see this point more clearly in the following protocol.

\textit{A Flawed Three-State Protocol.---}
Let us introduce a third dimension, $\ket{2}$, such that 
$\ket{0} $,  $\ket{1}  $, and $\ket{2}  $  form an orthonormal basis. We now consider a flawed three-state protocol wherein the states lie in three dimensions, rather than two. As before, our conjugate basis consists of the state $\ket{+}$. However, our computational basis consists of $\ket{0}$ and a flawed state, $\ket{1'}$. This flawed state takes the following form: 
\begin{equation}
    \ket{1'} = {\sqrt{1 - \epsilon}} \ket{1} + {\sqrt{ \epsilon} }
    \ket{2} .
    \label{eq:threestate}
\end{equation}

With some small amplitude, $\sqrt{\epsilon}$, $\ket{1'}$ lies in a third dimension, which is clearly a violation of the qubit assumption. Since the assumption is that Bob's measurement is either inaccessible or oblivious to this third dimension, Alice's inadvertent multi-dimensional modulation opens up a fatal attack strategy known as the unambiguous state discrimination (USD) attack \cite{Curras-Lorenzo2023,Ivanovic1987}. In this attack, Eve’s measurement either provides no information about the state or perfectly identifies it.

We now illustrate a potential USD attack for our flawed three-state protocol. Here, Eve performs a POVM with two elements. 
\begin{equation}
P_1 = \ket{0}\bra{0}+ \ket{1}\bra{1}
\end{equation}
and
\begin{equation}
P_2 = \ket{2}\bra{2} .
\end{equation}
Notice that the outcome of the second element allows Eve to identify the state $\ket{1'}$ unambiguously. We can also construct POVM measurements to identify the other two states unambiguously. Furthermore, Eve can exploit practical channel loss to mask her measurement. This is why state preparation imperfections which violate the qubit assumption are not loss-tolerant. Hence, the USD attack is a significant threat to the security of quantum protocols such as QKD. While we used a three-state protocol to illustrate the USD attack, it applies to any number of states.

\textit{Hidden Multi-Dimensional Modulation Side Channel.---} Now that we have reviewed how violating the qubit assumption can lead to the fatal USD attack, we can explain the hidden multi-dimensional modulation side channels which arise from this violation. We start by presenting a quantum optical model to elucidate the concept of these side channels, using our case example of time-varying polarization (BB84) encoding. Consider the single-photon component, $\ket{\psi^i_x}$, of an encoded quantum state with time-varying polarization, where $i \in \{0,1\}$ represents the bit choice and $x \in \{0,1\}$ represents the basis choice.

We first focus on the polarization degree of freedom. Note that active polarization encoding is typically achieved using an electro-optic phase modulator \cite{Gee1983} to modulate the phase along a specific polarization component. Without loss of generality, we consider an input (unmodulated) polarization state of $\frac{\ket{H}+\ket{V}}{\sqrt{2}}$. Here, $\ket{H}$ and $\ket{V}$ are orthogonal and linear polarization states, with $\ket{V}$ being the component along which phase modulation occurs. Hence, the polarization degree of freedom for a modulated state would take the following form: 

\begin{equation}
    \frac{\ket{H}+e^{i\phi^i_x}\ket{V}}{\sqrt{2}}
\end{equation}

As we consider BB84 encoding, $\phi^i_x$ would take on the following nominal values -- 0, $\pi$, $\pi/2$, $-\pi/2$. However, in practice, there are two factors to take note of. Firstly, the quantum optical state being modulated has a finite duration. Secondly, due to the bandwidth limitations of the equipment driving the electro-optic modulator, the applied phase change would not be constant over the duration of the quantum optical state. Hence, the modulation process involves a ``hidden'' temporal degree of freedom. 

We now incorporate this temporal degree of freedom into our model. Let $f(t)$ represent the intensity profile of the quantum optical state. Here, we assume a Gaussian intensity profile. Hence, the temporal probability distribution of the single-photon component would follow a Gaussian function, $f(t) = ae^{-\frac{t^2}{2c^2}}$. We choose $a$ such that $\int_{-\infty}^{\infty} f(t)=1$. The parameter $c$ is related to the full width at half maximum (FWHM) of the optical pulse ($2\sqrt{2\ln2}c$). Now, let $\phi^i_x(t)$ refer to the time-varying phase change applied by the phase modulator to the $\ket{V}$ component. We can then represent $\ket{\psi^{i}_{x}}$ as a summation of time-dependent polarization states over infinitesimal time-bins, $\ket{t}$, as follows (leaving out the global phase term, which we assume in this work to be time-independent):

\begin{equation}
    \ket{\psi^{i}_{x}} = \int_{-\infty}^{\infty} \sqrt{f(t)}  \biggl[ \ket{t} \otimes \frac{\ket{H}+e^{i\phi^i_x(t)}\ket{V}}{\sqrt{2}} \biggr] dt
    \label{eq:transmitted}
\end{equation}

It is clear that there is no separability between the time and polarization (encoding) degrees of freedom in \eqref{eq:transmitted}. In other words, the temporal degree of freedom is coupled to the encoding degree of freedom. We emphasize that while our model specifically considers polarization and time, these can easily be replaced with other degrees of freedom. In general, $\ket{H}$ and $\ket{V}$ in \eqref{eq:transmitted} can be replaced by $\ket{0}$ and $\ket{1}$, where $\ket{0}$ and $\ket{1}$ can be spatial modes, binary time bins, frequency bins, etc. Similarly, $\ket{t}$ in \eqref{eq:transmitted} can be a different degree of freedom. The hidden multi-dimensional modulation side channel persists as long as the encoding degree of freedom is coupled with another degree of freedom.

This additional degree of freedom increases the dimension of the Hilbert space spanned by the encoded states. As a result, the states may be linearly independent, opening up the possibility of an unambiguous state discrimination (USD) attack \cite{Duek2000,Ivanovic1987}. By exploiting channel loss to discard inconclusive USD measurements without detection, an eavesdropper can identify every state that reaches the detector.

The model presented in \eqref{eq:transmitted} will be used to compute the secret key rate when the time-varying encoding side channel is present. More specifically, the model will be used to compute the single-photon phase error rate (used to bound an eavesdropper's information on the secret key in a decoy state protocol \cite{Lo2005}).

We can intuitively understand the level of information leakage inherent in $\ket{\psi^{i}_{x}}$ by breaking down the state as follows \cite{Curras-Lorenzo2023}:

\begin{equation}
    \ket{\psi^i_x}=\sqrt{1-\epsilon^i_x} \ket{\theta^i_x} + \sqrt{\epsilon^i_x} \ket{\theta^{i \perp}_x}.
    \label{eq:decompose}
\end{equation}

Note the similarity between \eqref{eq:decompose} and \eqref{eq:threestate}. Here, $\ket{\theta^i_x}$ is the closest (in terms of fidelity) qubit state to $\ket{\psi^{i}_{x}}$. Here, we assume the form of $\ket{\theta^i_x}$ to match \eqref{eq:transmitted} with one key difference -- the polarization is not time-dependent (which results in separability between the temporal and polarization degrees of freedom). Hence, we can express $\ket{\theta^i_x}$ as  

\begin{equation}
    \ket{\theta^i_x} = \int_{-\infty}^{\infty} \sqrt{f(t)}  \biggl[ \ket{t} \otimes \frac{\ket{H}+e^{i\theta^i_x}\ket{V}}{\sqrt{2}} \biggr] dt
\end{equation} 

where $\theta^i_x$ represents a time-independent phase change applied by the phase modulator ($\theta^i_x$ is chosen such that $|\braket{\theta^i_x|\psi^i_x}|^2$ is maximal). The remaining component, $\ket{\theta^{i \perp}_x}$ is orthogonal to $\ket{\theta^i_x}$ and causes a violation of the qubit assumption by increasing the dimension of the Hilbert space. (Note, the implicit assumption of \eqref{eq:decompose} is that the qubit subspace in the larger Hilbert space is defined to be the polarization state with a constant temporal profile, i.e. {$\ket{\theta^i_x}$}.) The parameter $\epsilon^i_x$ represents the level of deviation of $\ket{\psi^i_x}$ from a qubit state due to time-varying encoding. It can be expressed as follows: 

\begin{equation}
    |\braket{\theta^i_x|\psi^i_x}|^2 = 1-\epsilon^i_x
    \label{eq:epsilon}
\end{equation} 

The impact of the side channel on security can be understood through the parameter $\epsilon^i_x$. We will later use this parameter in order to further understand and contextualize our secret key rate results. 

\textit{Impact of Time-Varying Encoding on the Secret Key Rate---}
Here, we concretely demonstrate the impact of a hidden multi-dimensional encoding side channel through key rate analysis, using our case example of time-varying polarization encoding. We employ a tight numerical key rate analysis technique based on semidefinite programming \cite{Primaatmaja2019} (see the Supplemental Material \footnote{See the Supplemental Material, which includes Refs. \cite{Ma2012_mdi_decoy,Bourassa2022,keysight,drve10mo,jdsu,scipy}, for the key rate analysis technique and methods for acquiring practical temporal phase profiles.} for further details). At a high level, this technique uses the Gram matrix of the transmitted states to bound Eve's information. Recall that here, the transmitted states can be expressed as per \eqref{eq:transmitted}, which we use to compute the inner products.

\begin{figure}[h!] 
    \centering
    \includegraphics[width=0.9\columnwidth]{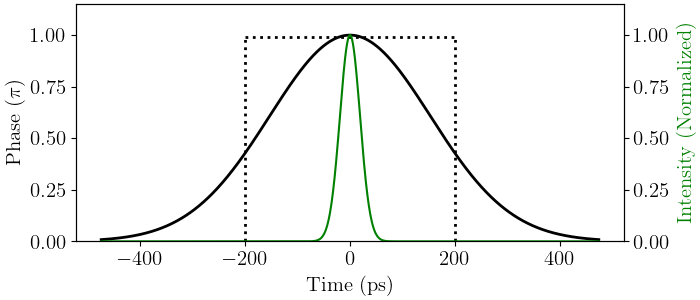}
    \caption{Normalized temporal phase profiles used in the key rate analysis (results shown in Figure \ref{fig:gaussian_keyrates}). The dashed black line indicates a square (time-invariant) profile while the solid black line indicates a Gaussian (time-varying) profile. Both profiles have a FWHM of 400 ps. The optical intensity profile, $f(t)$ (assumed in the key rate analysis) is also shown in green. It is a Gaussian intensity profile with a FWHM of 50 ps.}
    \label{fig:gaussian_profile}
\end{figure}

\begin{figure}[h!] 
    \centering
    \includegraphics[width=0.8\columnwidth]{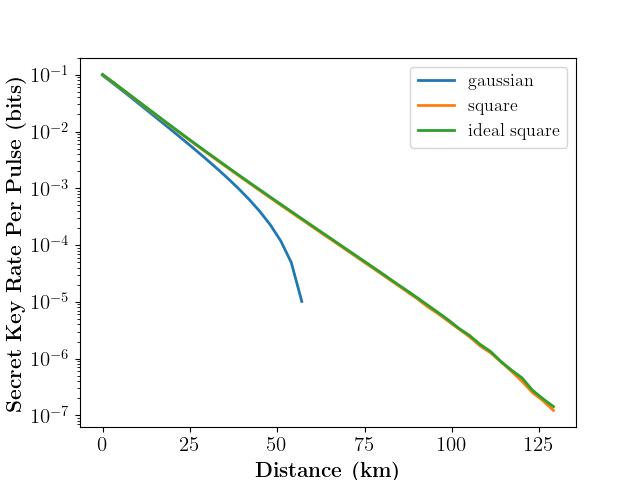}
    \caption{Key rate results (with respect to the distance between Alice/Bob and the central relay) for varying phase profiles. Blue: Gaussian phase profile (illustrated in Figure \ref{fig:gaussian_profile}). Orange: Square (time-independent over the duration of the optical pulse) phase profiles. The amplitude is set to be the average phase of the Gaussian phase profile (see (\ref{eq:average})). Green: Square phase profiles with ideal amplitudes.} 
    \label{fig:gaussian_keyrates}
\end{figure}

Using this technique, we simulate the secret key rate for two cases of the temporal phase profile, $\phi^i_x(t)$ -- time-independent (square) and Gaussian (see Figure \ref{fig:gaussian_profile}). The time-independent phase is set to be the average phase of the Gaussian phase profile. We define the average phase, $\gamma_{avg}$, of a phase profile, $\gamma(t)$ as follows:
    \begin{equation}
        \gamma_{avg} = \int_{-\infty}^{\infty} f(t) \gamma(t) dt.
        \label{eq:average}
    \end{equation}
Recall that $f(t)$ represents the Gaussian optical intensity profile of the transmitted states. We also simulate the secret key rate assuming a time-independent phase with a perfect BB84 amplitude (ideal square). The secret key rate results can be found in Figure \ref{fig:gaussian_keyrates}.

In Figure \ref{fig:gaussian_keyrates}, we see that the Gaussian phase profile leads to a significant (approximately 50\%) reduction in the maximum secure transmission distance, relative to the ideal case. Comparatively, despite having the same average encoding error as the Gaussian profile, the square profile only causes a marginal reduction in the maximum secure transmission distance, relative to the ideal case. Here comes a key conceptual point. A square phase profile implies that transmitted states satisfy the qubit assumption and hence do not carry the time-varying encoding side channel. This is assuming the optical intensity profile is negligible outside the modulation window, as is the case here. This results in a higher maximum secure transmission distance, as the protocol is now loss-tolerant \cite{Curras-Lorenzo2023}. In contrast, a temporally varying phase profile violates the qubit assumption. This provides additional dimensions to the Hilbert space and may enable a USD attack. Hence, the secure key rate is substantially reduced.

\textit{Practical Temporal Phase Profiles} --- Here, we present key rate analysis results using practical (experimental and simulated) $\phi^i_x(t)$. This allows us to concretely understand the impact of time-varying encoding in realistic QKD systems. We focus on high-speed phase modulation, as QKD systems operating at GHz clock rates are most impacted by bandwidth limitations, which lead to time-varying modulation (recall that a square modulation signal requires infinite bandwidth). High-speed transmission enhances the practicality of QKD, sparking recent interest in this area \cite{Roger2023, Boaron2018, Li2019, Grnenfelder2020, Yoshino2018, Woodward2021, Beutel2021, Chen2023, Lu2021}. Therefore, our study of this newly identified side channel is crucial.

We performed our experimental measurement of $\phi^i_x(t)$ using an electro-optic modulator of 15 GHz bandwidth, driven by an arbitrary waveform generator of 25 GHz bandwidth and amplifier of 11 GHz bandwidth. These bandwidths are typical of components used in today's GHz QKD experiments \cite{Yoshino2018,Lu2021,Kang2023,Roberts2018}. While limited bandwidth is a notable imperfection impacting the time-variance of phase modulation, other equipment-specific factors such as nonlinearity and the frequency response function also contribute. Measurement artifacts may also affect the results. This motivates us to consider optimistically simulated phase profiles as a generalized baseline. This provides us with a practical lower bound on the time-variance of the profiles. We consider a 50 ps wide optical pulse and a 200 ps wide phase modulation window, typically used at GHz clock rates \cite{Lu2021,Kang2023,Roberts2018,Yoshino2018,Li2019}. To focus on the impact of time-variance breaking the qubit assumption, we mitigate the influence of correlations, a separate side channel, by collecting $\phi^i_x(t)$ at a low repetition rate of 10 MHz. Refer to the Supplemental Material for details regarding the simulation and experimental methods. Figures 1 and 3 in the Supplemental Material provide the resulting simulated and experimental temporal phase profiles, respectively.

\begin{figure}[h]
    \centering
    \includegraphics[width=0.8\columnwidth]{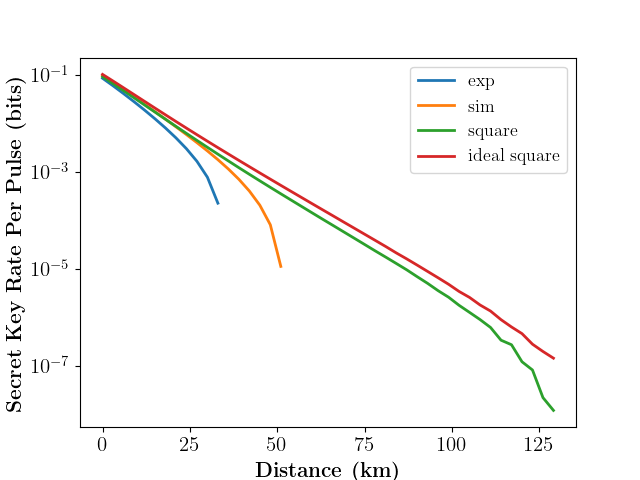}
    \caption{Key rate results (with respect to the distance between Alice/Bob and the central relay) for different phase profiles assuming an optical intensity profile with a 50 ps FWHM. The phase profiles have a 200 ps FWHM. Blue: Experimentally collected phase profiles. Orange: Simulated phase profile. Green: Square (time-independent over the duration of the optical pulse) phase profiles. The amplitude is set to be the average phase of the experimental phase profile (see (\ref{eq:average})). Red: Square phase profiles with ideal amplitudes.}
    \label{fig:keyrates}
\end{figure}

In Figure \ref{fig:keyrates}, we see that the experimentally collected phase profiles lead to a significant (approximately 70\%) reduction in the maximum secure transmission distance, relative to the ideal case. Since the simulated phase profiles are optimistic and do not consider any imperfections beyond bandwidth limitations, they perform slightly better. However, even in this relatively optimistic case, the maximum secure transmission distance is reduced by approximately 60\%. Figure \ref{fig:keyrates} also demonstrates the secret key rate when using square profiles with amplitudes corresponding to the average phase (see (\ref{eq:average})) of the experimental profiles (square). As the square profile satisfies the qubit assumption, it does not cause a drastic reduction in the maximum secure transmission distance.

To further understand our key rate results, we can compute the corresponding $\epsilon^i_x$ values for each case (see Table \ref{table1}). We can see that the experimentally collected and simulated phase profiles lead to $\epsilon^i_x$ values on the order of $10^{-3}$. Referring back to \cite{Curras-Lorenzo2023}, we can qualitatively understand the impact of $\epsilon^i_x$ of various orders of magnitude on the secret key rate. For a BB84 protocol using single photons, \cite{Curras-Lorenzo2023} reports a reduction in the maximum secure transmission distance of over 70\% (relative to the qubit case) for an $\epsilon^i_x$ on the order of $10^{-3}$. In addition to our key rate results, this further emphasizes the substantial impact which this side channel has on security.

\begin{table}[h!]
\centering
\caption{Table of $\epsilon^i_x$ values corresponding to the experimental and simulated cases in Figure \ref{fig:keyrates}.}
\begin{tabular}{|c||c|c|}
  \hline
   & $\pi$ & $\pi/2$ \\
  \hline\hline
  exp profile & $2.661\times10^{-3}$ & $8.193\times10^{-4}$ \\
  \hline
  sim profile & $1.019\times10^{-3}$ & $2.556\times10^{-4}$ \\
  \hline
\end{tabular}
\label{table1}
\end{table}

\begin{figure}[h]
    \centering
    \includegraphics[width=0.8\columnwidth]{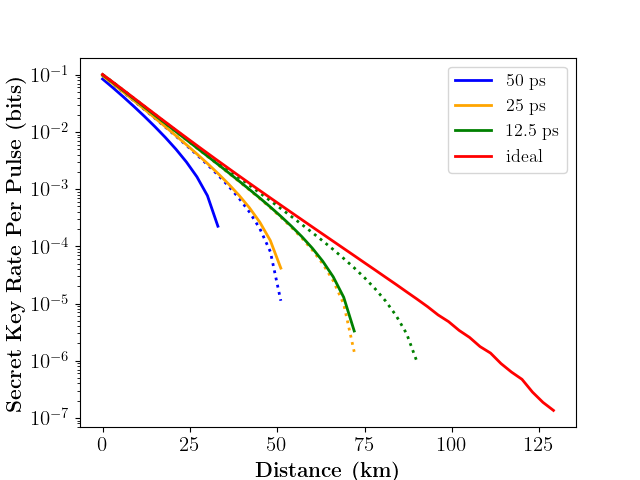}
    \caption{Key rate results (with respect to the distance between Alice/Bob and the central relay) using the experimentally collected phase profiles. The results from using the simulated phase profile are shown using a dotted line. Here, the FWHM of the Gaussian optical intensity profile is varied. Blue: 50 ps width. Orange: 25 ps width. Green: 12.5 ps width. Red: Square phase profiles with ideal amplitudes.}
    \label{fig:keyrates_pw}
\end{figure}

Notice that the smaller the pulse width, the more constant the temporal phase profile would be over the pulse duration, leading to the side channel having a smaller impact. This is confirmed by Figure \ref{fig:keyrates_pw}, showing the simulated secret key rates for optical pulse widths (FWHM) of 25 ps and 12.5 ps.

The corresponding $\epsilon^i_x$ values are also provided in Table \ref{table2}. As expected, we see dramatic improvements in the secret key rate as the optical pulse width gets smaller in relation to the 200 ps modulation pulse width (along with order of magnitude changes in $\epsilon^i_x$). This suggests that this side channel can be mitigated by either shortening the optical pulse width or increasing the modulation pulse width. Increasing the modulation pulse width would increase the duty cycle of the modulation pulse train - this would exacerbate the issue of pulse correlations, which are a major security concern in high clock rate QKD systems \cite{Pereira2020}. Shortening the optical pulse width is a relatively viable option -- however, for substantial improvements to the secret key rate, techniques such as mode-locking are required for generating pulses on the order of ps \cite{Martinez2011,Zhao2021}.

\begin{table}[h]
\centering
\caption{Table of $\epsilon^i_x$ values corresponding to the experimental cases in Figure \ref{fig:keyrates_pw} (varying optical pulse widths).}
\begin{tabular}{|c||c|c|}
  \hline
   & $\pi$ & $\pi/2$ \\
  \hline\hline
  50 ps & $2.661\times10^{-3}$ & $8.193\times10^{-4}$ \\
  \hline
  25 ps & $8.221\times10^{-4}$ & $1.684\times10^{-4}$ \\
  \hline
  12.5 ps & $1.141\times10^{-4}$ &  $4.120\times10^{-5}$ \\
  \hline
\end{tabular}
\label{table2}
\end{table}

\textit{Conclusion.---} In this work, we have identified a new class of side channels which affect quantum cryptographic protocols -- in practical experiments, the state preparation process may be entangled with hidden degrees of freedom. We show that this would violate the qubit assumption, opening up a USD attack and severely limiting the maximum secure transmission distance. As an illustrative example, we consider time-varying polarization encoding in GHz clock rate QKD sources exploying active modulation. Through this example, our experimental and simulated side channel characterization revealed that this side channel can reduce the secret key rate by several orders of magnitude, as shown in Figure \ref{fig:keyrates}. Although we identified a potential mitigation method for our illustrative example, the optimal solution would be to eliminate the time degree of freedom from the encoding process entirely, ensuring the qubit assumption is not violated. At GHz repetition rates, this would be practically unfeasible when using active modulation due to bandwidth limitations. Hence, this work provides further motivation for the use of passive QKD sources \cite{Wang2023,Zapatero2023,Hu2023}. Crucially, our work demonstrates how seemingly minor practical factors in experimentally realized quantum protocols can lead to substantial security risks. This highlights the need for experimentalists to be highly aware of imperfections in experiments.

\begin{acknowledgments}
We would like to thank Reem Mandil, Alexander Greenwood, and Andi Shahaj for their support. This research was supported in part by NSERC, CFI, ORF, MITACS, and DRDC.
\end{acknowledgments}

\bibliographystyle{apsrev4-1}
\bibliography{biblio}

\widetext
\clearpage

\begin{center}
    \textbf{Supplemental Material}
\end{center}

\setcounter{equation}{0}
\setcounter{figure}{0}
\setcounter{table}{0}
\setcounter{page}{1}

\makeatletter
\renewcommand{\theequation}{S\arabic{equation}}
\renewcommand{\thefigure}{S\arabic{figure}}
\renewcommand{\bibnumfmt}[1]{[S#1]}
\renewcommand{\citenumfont}[1]{S#1}

\section{Key Rate Analysis Technique}

\label{sec:technique}
A symmetric polarization encoding (BB84) decoy state MDI QKD protocol with intensity settings of 0.05, 0.1, and 0.6 (near typical values \cite{Ma2012_mdi_decoy}) was assumed in the key rate analysis. Without loss of generality, we consider a projection onto the $\ket{\Psi^+}$ Bell state as a passing measurement. A fiber loss of 0.2 dB/km and a dark count rate of $10^{-6}$/pulse was assumed. We assume that the temporal encoding profiles of Bob's states are the same as the temporal encoding profiles of Alice's states. 

Now, we provide a high level picture for the key rate simulation technique \cite{Primaatmaja2019} which we employ. This technique bounds the secret key rate by bounding the single photon phase error rate ($e_{ph}$) of the QKD protocol using numerical optimization. The optimization problem can be written as follows.
  
\begingroup
\setlength{\tabcolsep}{4pt} 
\renewcommand{\arraystretch}{2} 
\begin{center}
\begin{tabular}{ l l }
\texttt{maximize} & $e_{ph}$  \\ 
\texttt{s.t.}  & $\braket{\psi^{i'}_{x'}\varphi^{j'}_{y'}|\psi^{i}_{x}\varphi^{j}_{y}}_{AB} = \sum_{z}\braket{e^{i',j'}_{x',y',z}|e^{i,j}_{x,y,z}}_E$ (a)\\
 &     $p_{\texttt{pass},L}^{i,j,x,y} \leq \braket{e^{i,j}_{x,y,P}|e^{i,j}_{x,y,P}}_E \leq p_{\texttt{pass},U}^{i,j,x,y}$ (b)\\
 & $0 \leq e_{ph} \leq \frac{1}{2}$ (c)\\
 & $G_E\succeq 0$ (d)
\end{tabular}
\end{center}
\endgroup

Here,
\begin{itemize}
    \item $\ket{\psi^i_x}_A$ and $\ket{\phi^j_y}_B$ represent the single photon component of Alice and Bob's transmitted coherent states, respectively. Superscripts indicate the basis choice (0 denoting the key generation basis and 1 denoting the conjugate basis) while subscripts indicate the encoded bit (0 or 1). Hence, $\braket{\psi^{i'}_{x'}\varphi^{j'}_{y'}|\psi^{i}_{x}\varphi^{j}_{y}}_{AB}$ represent the elements of Alice and Bob's Gramian matrix \cite{Primaatmaja2019}. It is through these inner products that the temporal phase profiles enter the key rate analysis. Note that $\ket{\psi^i_x}_A$ and $\ket{\phi^j_y}_B$ take the form given by Equation 6 in the main paper. 
    \item $\ket{e_{x,y,z}^{i,j}}_E$ (Eve's state) can be defined as per Equation \ref{eq:evol_to_Eve}, through the unitary evolution of Alice and Bob's transmitted state. Here, $z \in \{P,F\}$ represents Eve's announcement (pass/fail) in the protocol and $\braket{z|z'}_Z=\delta_{z,z'}$.  Note that $\ket{e^{i,j}_{x,y,z}}_E$ are subnormalized such that $\ket{\psi^{i}_{x}\varphi^{j}_{y}}_{AB}$ are normalized.
    \begin{equation}\label{eq:evol_to_Eve}
    \ket{\psi^{i}_{x}\varphi^{j}_{y}}_{AB} \rightarrow \sum_{z=P,F}\ket{e^{i,j}_{x,y,z}}_E\ket{z}_Z
    \end{equation}
    \item From Equation \ref{eq:evol_to_Eve}, it is apparent that $\braket{e^{i,j}_{x,y,P}|e^{i,j}_{x,y,P}}_E$ refers to the passing probability of the transmitted state $\ket{\psi^{i}_{x}\varphi^{j}_{y}}_{AB}$, since $\ket{e^{i,j}_{x,y,z}}_E$ are subnormalized. The variables $p_{\texttt{pass},L}^{i,j,x,y}$ and $p_{\texttt{pass},U}^{i,j,x,y}$ refer to the upper and lower bound on this passing probability. They can be found through the passing probability of Alice and Bob's overall transmitted coherent state \cite{Bourassa2022}. 
    \item $e_{ph}$ can be expressed as follows using the elements of Eve's Gramian matrix \cite{Bourassa2022}.
    \begin{equation}
    \frac{1}{2} - \frac{Re\left(\braket{e^{0,0}_{0,0,P}|e^{0,0}_{1,1,P}}_E+\braket{e^{0,0}_{0,1,P}|e^{0,0}_{1,0,P}}_E\right)}{\sum_{x,y} p_{\texttt{pass},L}^{0,0,x,y}}
    \end{equation}
    \item $G_E$ refers to Eve's Gramian matrix (which has elements $\braket{e^{i',j'}_{x',y',z'}|e^{i,j}_{x,y,z}}_E$).
\end{itemize}

Overall, this is an optimization of $e_{ph}$ with respect to Eve's Gramian matrix. The known values in this optimization problem are Alice and Bob's transmitted states. The unknown values are $e_{ph}$ (the objective function) and Eve's Gramian matrix, $G_E$.  

After bounding $e_{ph}$, we can evaluate the secret key rate, $R$. The formula for the asymptotic secret key rate in decoy state MDI QKD is as follows (neglecting the effects of sifting and assuming perfect error correction efficiency) \cite{Ma2012_mdi_decoy}.  

\begin{equation}\label{eq:KR_formula}
    R \geq  p_{\texttt{pass},L}^{0,0}[1-h_2(e_{ph})] - Q h_2(E_{bit}),
\end{equation}

Here, 

\begin{itemize}
    \item $p_{\texttt{pass},L}^{0,0} = \sum_{x,y} p_{\texttt{pass},L}^{0,0,x,y} $ refers to the total passing probability (lower bound) of the transmitted states (single photon component) in the key generation basis ($i,j=0$) over all possible $x,y$.
    \item $Q$ refers to the passing probability of signal states in the key generation basis.
    \item $E_{bit}$ is the bit error rate. Concretely, it's the proportion of passed signal states in the key generation basis in which Alice and Bob encode opposite bits. 
    \item $h_2$ refers to the binary entropy function. 
\end{itemize}

\section{Experimental and Simulated Characterization of Temporal Phase Profiles}
Here, we describe how we practically acquire the time-varying phase profiles, $\phi^i_x(t)$, from Equation 6 in the main paper. We acquire the phase profiles using two methods -- experiment and simulation. The corresponding key rate results are discussed in the \textit{Practical Temporal Phase Profiles} section in the main paper. 

We start by reviewing electro-optic phase modulation \cite{Gee1983}. An electro-optic phase modulator works through a traveling wave mechanism in which an optical signal and traveling RF (radiofrequency) signal propagate together through the modulator with their speeds matched. The voltage required to create a $\pi$ phase change is referred to as $V_\pi$ (more generally, we will use $V_{\phi}$ to refer to the voltage required to create a $\phi$ phase change). 

As we consider BB84 encoding, we are interested in the temporal phase profiles resulting from the following phase modulations - $\pi/2$, $\pi$, and $-\pi/2$. The width of the phase modulation window (i.e. the width of the voltage pulse used to modulate phase) must be adequately larger than the width of the optical pulse being modulated. Here, we consider a 50 ps wide optical pulse and a 200 ps wide phase modulation window, typically used at GHz clock rates \cite{Lu2021,Kang2023,Roberts2018,Yoshino2018,Li2019}. Ideally, the voltage pulse used to modulate phase should be square, resulting in time-invariant modulation over the duration of the optical pulse. However, the bandwidth limitations of the arbitrary waveform generator (AWG) and RF amplifier (which is often necessary to create voltage pulses with the required amplitude) result in distorted voltage pulses.

\begin{figure}[h!]
    \centering
    \includegraphics[width=0.4\columnwidth]{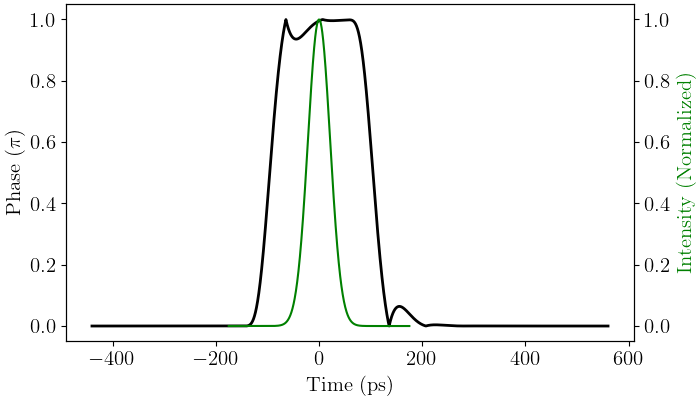}
    \caption{Simulated $\pi$ phase profile (the simulated $\pm \pi/2$ phase profiles are simply a scaled version of this). This was generated by applying Butterworth filters to a 200 ps wide square pulse. The Butterworth filters represent the AWG, amplifier, and electro-optic modulator. The optical pulse intensity profile assumed in this work (50 ps FWHM Gaussian pulse) is also shown in green.}
    \label{fig:sim_profile}
\end{figure}

While limited bandwidth is a notable imperfection impacting the time-variance of phase modulation, other equipment-specific factors, such as nonlinearity and the frequency response function, may also contribute. This motivates us to provide optimistically simulated phase profiles in addition to our measured phase profiles as a generalized baseline. The simulated phase profiles solely consider the bandwidth limitations of the AWG, RF amplifier, and electro-optic modulator. We optimistically approximate the frequency response of each component using a simple Butterworth filter of order 2, with -3 dB points of 25 GHz, 11 GHz, and 15 GHz, respectively (corresponding to the equipment specifications). Butterworth filters are an optimistic choice, as their frequency response in the passband is relatively flat. Furthermore, at an order of 2, the frequency response falls off relatively gradually past the passband (in comparison to the frequency response of our state-of-the-art equipment) \cite{keysight,drve10mo,jdsu}. These filters were created and applied to a 200 ps wide square pulse in order to generate the simulated profile (using the SciPy signal processing class) \cite{scipy}. The result can be seen in Figure \ref{fig:sim_profile}. 

\newpage

\begin{figure}[h!]
    \centering
    \includegraphics[width=0.7\columnwidth]{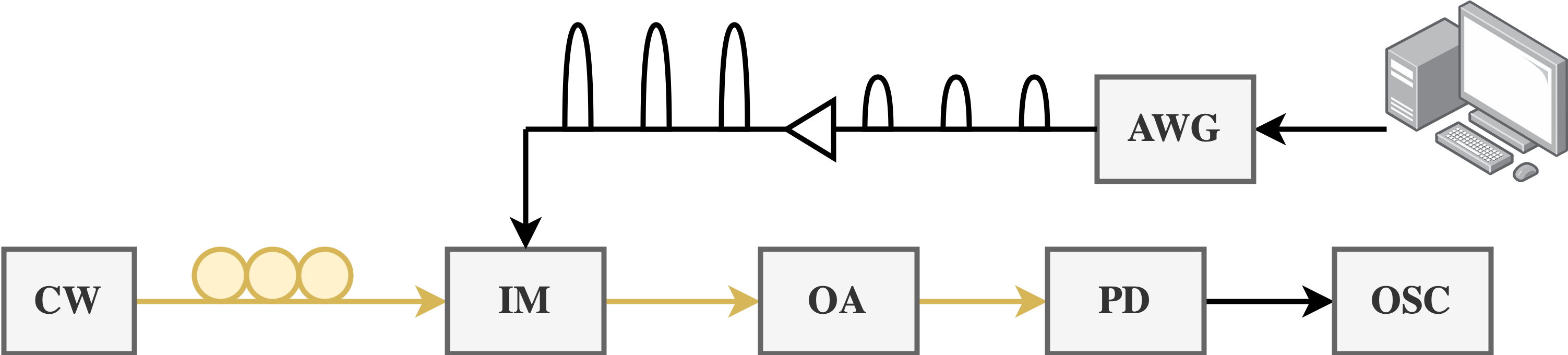}
    \caption{Experimental setup for generating and measuring phase profiles. The 25 GHz Keysight M8195A AWG and 11 GHz ixBlue DR-VE-10-MO RF amplifier are used to generated voltage pulses which are sent to the JDSU OC-192 intensity modulator (IM). This intensity modulator is a phase modulator with a built in Mach-Zehnder interferometer to convert phase changes to intensity changes. The optical intensity profile (related to the optical phase profile through Equation \ref{phase-voltage}) is collected using the Newport D-8ir 8 ps photodiode (PD) and the 12 GHz Agilent DSO81204A oscilloscope (OSC) after being amplified by a Thorlabs semiconductor optical amplifier (OA). The OA and PD were operated within their linear response region. Yellow indicates an optical fiber connection.}
    \label{fig:setup2}
\end{figure}

We now move to our experimentally acquired temporal phase profiles.  The experimental setup for generating and measuring the phase profiles is shown in Figure \ref{fig:setup2}. The 25 GHz AWG was set to generate 200 ps wide (FWHM) square voltage pulses (the pulse repetition frequency was set to 10 MHz to reduce pulse correlations down to negligible levels). This pulse train was then amplified by an 11 GHz amplifier so that the voltage pulses have the necessary amplitude to generate the desired phase change. Using this setup, voltage pulses with an amplitude of $V_{\pi}$, $V_{\pi/2}$, and $V_{-\pi/2}$ were generated in order to acquire the $\pi$, $\pi/2$, and $-\pi/2$ temporal phase profiles. For our electro-optic modulator, $V_{\pi}$, $V_{\pi/2}$, and $V_{-\pi/2}$ are approximately 5.5 V, 2.75 V, and -2.75 V, respectively. 

The $\pi$, $\pi/2$, and $-\pi/2$ temporal phase profiles were each acquired by using the corresponding amplified voltage pulses to drive an electro-optic intensity modulator.  An electro-optic intensity modulator uses electro-optic phase modulation in conjunction with Mach-Zehnder interferometry in order to modulate intensity. The phase change ($\phi$) applied by the intensity modulator is related to the normalized output intensity ($I$) through Equation \ref{phase-voltage} when the intensity modulator is biased at a minimum. This relationship can be used to determine a phase profile from its corresponding intensity profile. 

\begin{equation}
    \phi = 2cos^{-1}(\sqrt{1-I})
    \label{phase-voltage}
\end{equation}

\begin{figure}[h!]
    \centering
    \includegraphics[width=0.8\columnwidth]{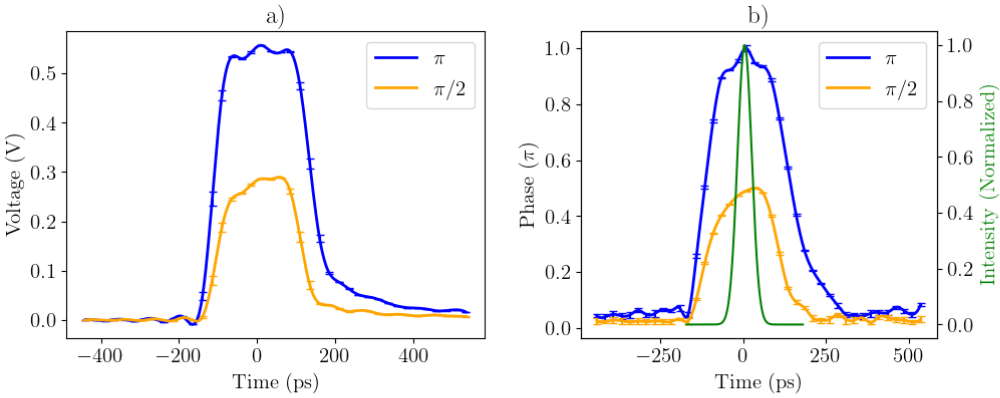}
    \caption{a),b) Voltage pulses used for polarization/phase modulation as well as the corresponding phase profiles. Cubic spline interpolations \cite{scipy} of the profiles are shown using a solid line. The optical pulse intensity profile assumed in this work (50 ps FWHM Gaussian pulse) is also shown in green. Each profile was measured 10 times over an extended time period of several minutes (each time, oscilloscope averaging was used and set to 1024 to average out random noise, such as electronic noise) - the error bars represent the standard deviation in the 10 trials.}
    \label{fig:profiles}
\end{figure}

The resulting temporal phase profiles are shown in Figure \ref{fig:profiles} (excluding the $-\pi$/2 profile, which we found to be nearly identical to the $\pi/2$ profile in shape). The corresponding voltage profiles are also shown. Evidently, the $\pi$ and $\pi/2$ phase profiles have a significant level of temporal variation (especially in comparison to the optimistically simulated phase profile from Figure \ref{fig:sim_profile}, as expected). 

\end{document}